\documentclass[conference,10pt,letterpaper]{IEEEtran}

\usepackage{cite}
\usepackage{epsfig}
\usepackage{epstopdf}
\usepackage{graphicx}
\usepackage[dvipsnames]{xcolor}
\usepackage{tikz}
\usetikzlibrary{arrows,positioning,shapes.geometric,circuits.logic.US}
\usepackage{pgfplots}
\usepackage{multirow}
\usepackage{upgreek}
\usepackage{amssymb}
\usepackage{amsmath}
\usepackage{cases}
\usepackage{url}
\usepackage{pifont}
\usepackage{bm}
\usepackage{amsthm}

\usepackage{tabularx}
\usepackage{booktabs}
\usepackage{filecontents}
\usepackage{subcaption}
\newcolumntype{Y}{>{\centering\arraybackslash}X}

\DeclareMathOperator*{\argmax}{arg\,max}
\theoremstyle{definition} 
\theoremstyle{definition} 
\theoremstyle{definition} 
\theoremstyle{definition}

\newcolumntype{M}[1]{>{\centering\arraybackslash}m{#1}}

\usepackage[linesnumbered, ruled, vlined, noend]{algorithm2e}

\usepackage{algorithmic}

\usepackage{array}

\newcommand{\fixme}[2]{\ifx&#2&{\leavevmode\color{red}#1}\else{\leavevmode\color{red}FIXME\{}#1{\leavevmode\color{red}\}}\footnote{{\leavevmode\color{red}#2}}\PackageWarning{Fixme}{#1: #2}\fi}

\newcommand{\newstuff}[2]{\ifx&#2&{\leavevmode\color{blue}#1}\else{\leavevmode\color{blue}FIXME\{}#1{\leavevmode\color{blue}\}}\footnote{{\leavevmode\color{blue}#2}}\PackageWarning{Newstuff}{#1: #2}\fi}

\title{High-performance low-complexity error pattern generation for ORBGRAND decoding}

\author{Carlo Condo, Valerio Bioglio, Ingmar Land}

\begin{document}

\maketitle
\begin{abstract}
Guessing Random Additive Noise Decoding (GRAND) is a recently proposed decoding method searching for the error pattern applied to the transmitted codeword.
Ordered reliability bit GRAND (ORBGRAND) uses soft channel information to reorder entries of error patterns, generating them according to a fixed schedule, i.e. their logistic weight. 
In this paper, we show that every good ORBGRAND scheduling should follow an universal partial order, and we present an algorithm to generate the logistic weight order accordingly. 
We then propose an improved error pattern schedule that can improve the performance of ORBGRAND of 0.5dB at a block error rate (BLER) of $10^{-5}$, with increasing gains as the BLER decreases. 
This schedule can be closely approximated with a low-complexity generation algorithm that is shown to incur no BLER degradation.

\end{abstract}

\begin{IEEEkeywords}
GRAND, ORBGRAND, short codes, polar codes, BCH codes, ML decoding
\end{IEEEkeywords}

\IEEEpeerreviewmaketitle

\section{Introduction}\label{sec:intro}

Maximum Likelihood (ML) decoding can be seen as a universal decoding scheme for any error-correcting code. 
A binary code $\mathcal{C}$ of length $N$ and dimension $K$ allows to map a message $m$, composed of a string of $K$ bits, into a codeword $x$ of $N$ bits to be transmitted over a noisy channel.  
The channel alters the transmitted codeword such that the receiver obtains a vector $y$ composed by $N$ symbols. 
In theory, an ML decoder should compare $y$ with all the $2^k$ codewords in the codebook generated by the error-correcting code, and select the one closest to $y$, namely the candidate for which the error probability is minimized. 
More formally, given a codebook $\mathcal{C}$, an ML decoder finds the codeword candidate $\hat{x}$ such that
\begin{equation}
\label{eq:ML_def}
\hat{x} = \argmax_{x \in \mathcal{C}} p(y|x).
\end{equation} 
Even if ML decoding is optimal, its high complexity makes it impractical for the decoding of many codes. 
Recently, research on decoding latency reduction of near-ML decoders has been gaining momentum \cite{near_ML_overview}. 

An example of a near-ML decoder is given by the ordered statistics decoding (OSD) algorithm \cite{osd_paper}. 
This algorithm performs the Gauss-Jordan elimination procedure on the matrix obtained by extracting the columns of the generator matrix corresponding to the most reliable received symbols. 
Next, all the error patterns of Hamming weight not larger than a provided value $h$ are tested in order to determine the best candidate. 
The decoding latency of OSD is quite large due to the sorting of the symbols and the computation of the Gaussian elimination; moreover, the number of error patterns dramatically grows with $h$. 
A wise choice of the parameters, however, makes OSD tractable, significantly reducing the decoding latency compared to a full ML decoder. 

Recently, Guessing Random Additive Noise Decoders (GRAND) \cite{GRAND_first} have been proposed as an alternative to OSD to perform ML decoding or near-ML decoding with limited complexity. 
GRAND is a universal decoder, i.e., it can be used to decode any code. 
The algorithm checks all error patterns, and it schedules them in descending likelihood order, given the channel model employed. 
GRAND with abandonment (GRANDAB) has been proposed in \cite{GRAND_first} to reduce the decoding latency by limiting the number of error patterns to test. 
While GRAND and GRANDAB target channels without reliability information, Symbol Reliability GRAND (SRGRAND) \cite{SGRAND_first,soft_GRAND} uses the hard decisions plus one-bit reliability information to improve decoding accuracy. 
Ordered Reliability Bit GRAND (ORBGRAND) \cite{ORBGRAND_first} operates on soft channel outputs and uses logistic weights to schedule the error patterns after sorting the received symbols in order of reliability. 

One crucial issue for practical implementation of GRAND decoders is the need to store and access the error patterns, whose number can be very large.
ORBGRAND is one of the most promising embodiments of GRAND in this sense, since it couples an algorithmic error pattern schedule with soft information, leading to good performance with a limited number of test patterns.
The authors in \cite{ORBGRAND_arch} generate the error patterns on-the-fly, though limiting themselves to Hamming weights not larger than $h=3$.
While this method allows to reduce the hardware complexity of the decoder, it limits its error-correction performance \cite{ORBGRAND_first}. 


In this paper, we show how the order of error patterns \cite{ORBGRAND_first} to be scheduled is connected to the Universal Partial Order (UPO) proposed in \cite{UPO_construction} for the design of polar codes, and show that the ORBGRAND error pattern sequence can be created on-the-fly following the UPO properties. 
We then propose an alternative schedule, of comparable generation complexity, that can substantially improve the decoding performance of ORBGRAND; it yields gains of up to 0.5dB at a block error rate (BLER) of $10^{-5}$, and can match the performance of the original schedule with an order of magnitude fewer error patterns and thus lower complexity.
Finally, a low-complexity algorithm for an on-the-fly approximated implementation of the new schedule is presented, and it is shown not to incur any performance degradation down to a BLER of $10^{-7}$.

\section{Preliminaries}\label{sec:pre}
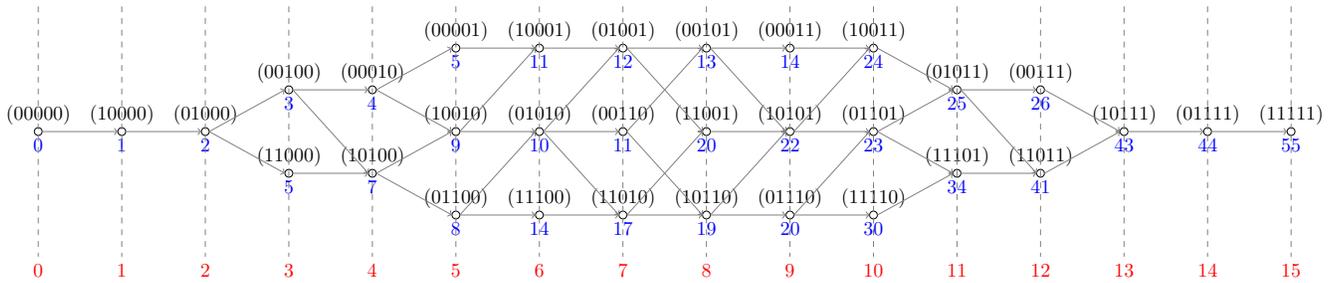
\begin{figure*}[tb]
	\begin{center}
		\resizebox{.98\textwidth}{!}{\begin{tikzpicture}

\def\ra{1.5}
\def\arr{.07}

\draw  ({0*\ra},{0*\ra}) ellipse ({\arr} and {\arr}) node[above] {$(00000)$} node[below,text=blue] {$0$};
\draw  ({1*\ra},{0*\ra}) ellipse ({\arr} and {\arr}) node[above] {$(10000)$} node[below,text=blue] {$1$};
\draw  ({2*\ra},{0*\ra}) ellipse ({\arr} and {\arr}) node[above] {$(01000)$} node[below,text=blue] {$2$};
\draw  ({3*\ra},{.5*\ra}) ellipse ({\arr} and {\arr}) node[above] {$(00100)$} node[below,text=blue] {$3$};
\draw  ({3*\ra},{-.5*\ra}) ellipse ({\arr} and {\arr}) node[above] {$(11000)$} node[below,text=blue] {$5$};
\draw  ({4*\ra},{.5*\ra}) ellipse ({\arr} and {\arr}) node[above] {$(00010)$} node[below,text=blue] {$4$};
\draw  ({4*\ra},{-.5*\ra}) ellipse({\arr} and {\arr}) node[above] {$(10100)$} node[below,text=blue] {$7$};
\draw  ({5*\ra},{1*\ra}) ellipse ({\arr} and {\arr}) node[above] {$(00001)$} node[below,text=blue] {$5$};
\draw  ({5*\ra},{0*\ra}) ellipse ({\arr} and {\arr}) node[above] {$(10010)$} node[below,text=blue] {$9$};
\draw  ({5*\ra},{-1*\ra}) ellipse ({\arr} and {\arr}) node[above] {$(01100)$} node[below,text=blue] {$8$};
\draw  ({6*\ra},{1*\ra}) ellipse ({\arr} and {\arr}) node[above] {$(10001)$} node[below,text=blue] {$11$};
\draw  ({6*\ra},{0*\ra}) ellipse ({\arr} and {\arr}) node[above] {$(01010)$} node[below,text=blue] {$10$};
\draw  ({6*\ra},{-1*\ra}) ellipse ({\arr} and {\arr}) node[above] {$(11100)$} node[below,text=blue] {$14$};
\draw  ({7*\ra},{1*\ra}) ellipse ({\arr} and {\arr}) node[above] {$(01001)$} node[below,text=blue] {$12$};
\draw  ({7*\ra},{0*\ra}) ellipse ({\arr} and {\arr}) node[above] {$(00110)$} node[below,text=blue] {$11$};
\draw  ({7*\ra},{-1*\ra}) ellipse ({\arr} and {\arr}) node[above] {$(11010)$} node[below,text=blue] {$17$};
\draw  ({8*\ra},{1*\ra}) ellipse ({\arr} and {\arr}) node[above] {$(00101)$} node[below,text=blue] {$13$};
\draw  ({8*\ra},{0*\ra}) ellipse ({\arr} and {\arr}) node[above] {$(11001)$} node[below,text=blue] {$20$};
\draw  ({8*\ra},{-1*\ra}) ellipse ({\arr} and {\arr}) node[above] {$(10110)$} node[below,text=blue] {$19$};
\draw  ({9*\ra},{1*\ra}) ellipse ({\arr} and {\arr}) node[above] {$(00011)$} node[below,text=blue] {$14$};
\draw  ({9*\ra},{0*\ra}) ellipse ({\arr} and {\arr}) node[above] {$(10101)$} node[below,text=blue] {$22$};
\draw  ({9*\ra},{-1*\ra}) ellipse ({\arr} and {\arr}) node[above] {$(01110)$} node[below,text=blue] {$20$};
\draw  ({10*\ra},{1*\ra}) ellipse ({\arr} and {\arr}) node[above] {$(10011)$} node[below,text=blue] {$24$};
\draw  ({10*\ra},{0*\ra}) ellipse ({\arr} and {\arr}) node[above] {$(01101)$} node[below,text=blue] {$23$};
\draw  ({10*\ra},{-1*\ra}) ellipse ({\arr} and {\arr}) node[above] {$(11110)$} node[below,text=blue] {$30$};
\draw  ({11*\ra},{.5*\ra}) ellipse ({\arr} and {\arr}) node[above] {$(01011)$} node[below,text=blue] {$25$};
\draw  ({11*\ra},{-.5*\ra}) ellipse ({\arr} and {\arr}) node[above] {$(11101)$} node[below,text=blue] {$34$};
\draw  ({12*\ra},{.5*\ra}) ellipse ({\arr} and {\arr}) node[above] {$(00111)$} node[below,text=blue] {$26$};
\draw  ({12*\ra},{-.5*\ra}) ellipse ({\arr} and {\arr}) node[above] {$(11011)$} node[below,text=blue] {$41$};
\draw  ({13*\ra},{0*\ra}) ellipse ({\arr} and {\arr}) node[above] {$(10111)$} node[below,text=blue] {$43$};
\draw  ({14*\ra},{0*\ra}) ellipse ({\arr} and {\arr}) node[above] {$(01111)$} node[below,text=blue] {$44$};
\draw  ({15*\ra},{0*\ra}) ellipse ({\arr} and {\arr}) node[above] {$(11111)$} node[below,text=blue] {$55$};

\draw[->,gray] ({0*\ra+\arr},{0*\ra}) -- ({1*\ra-\arr},{0*\ra});
\draw[->,gray] ({1*\ra+\arr},{0*\ra}) -- ({2*\ra-\arr},{0*\ra});
\draw[->,gray] ({2*\ra+\arr},{0*\ra}) -- ({3*\ra-\arr},{.5*\ra});
\draw[->,gray] ({2*\ra+\arr},{0*\ra}) -- ({3*\ra-\arr},{-.5*\ra});
\draw[->,gray] ({3*\ra+\arr},{.5*\ra}) -- ({4*\ra-\arr},{.5*\ra});
\draw[->,gray] ({3*\ra+\arr},{.5*\ra}) -- ({4*\ra-\arr},{-.5*\ra});
\draw[->,gray] ({3*\ra+\arr},{-.5*\ra}) -- ({4*\ra-\arr},{-.5*\ra});
\draw[->,gray] ({4*\ra+\arr},{.5*\ra}) -- ({5*\ra-\arr},{1*\ra});
\draw[->,gray] ({4*\ra+\arr},{.5*\ra}) -- ({5*\ra-\arr},{0*\ra});
\draw[->,gray] ({4*\ra+\arr},{-.5*\ra}) -- ({5*\ra-\arr},{0*\ra});
\draw[->,gray] ({4*\ra+\arr},{-.5*\ra}) -- ({5*\ra-\arr},{-1*\ra});
\draw[->,gray] ({5*\ra+\arr},{1*\ra}) -- ({6*\ra-\arr},{1*\ra});
\draw[->,gray] ({5*\ra+\arr},{0*\ra}) -- ({6*\ra-\arr},{1*\ra});
\draw[->,gray] ({5*\ra+\arr},{0*\ra}) -- ({6*\ra-\arr},{0*\ra});
\draw[->,gray] ({5*\ra+\arr},{-1*\ra}) -- ({6*\ra-\arr},{0*\ra});
\draw[->,gray] ({5*\ra+\arr},{-1*\ra}) -- ({6*\ra-\arr},{-1*\ra});
\draw[->,gray] ({6*\ra+\arr},{1*\ra}) -- ({7*\ra-\arr},{1*\ra});
\draw[->,gray] ({6*\ra+\arr},{0*\ra}) -- ({7*\ra-\arr},{1*\ra});
\draw[->,gray] ({6*\ra+\arr},{0*\ra}) -- ({7*\ra-\arr},{0*\ra});
\draw[->,gray] ({6*\ra+\arr},{0*\ra}) -- ({7*\ra-\arr},{-1*\ra});
\draw[->,gray] ({6*\ra+\arr},{-1*\ra}) -- ({7*\ra-\arr},{-1*\ra});
\draw[->,gray] ({7*\ra+\arr},{1*\ra}) -- ({8*\ra-\arr},{1*\ra});
\draw[->,gray] ({7*\ra+\arr},{1*\ra}) -- ({8*\ra-\arr},{0*\ra});
\draw[->,gray] ({7*\ra+\arr},{0*\ra}) -- ({8*\ra-\arr},{1*\ra});
\draw[->,gray] ({7*\ra+\arr},{0*\ra}) -- ({8*\ra-\arr},{-1*\ra});
\draw[->,gray] ({7*\ra+\arr},{-1*\ra}) -- ({8*\ra-\arr},{0*\ra});
\draw[->,gray] ({7*\ra+\arr},{-1*\ra}) -- ({8*\ra-\arr},{-1*\ra});
\draw[->,gray] ({8*\ra+\arr},{1*\ra}) -- ({9*\ra-\arr},{1*\ra});
\draw[->,gray] ({8*\ra+\arr},{1*\ra}) -- ({9*\ra-\arr},{0*\ra});
\draw[->,gray] ({8*\ra+\arr},{0*\ra}) -- ({9*\ra-\arr},{0*\ra});
\draw[->,gray] ({8*\ra+\arr},{-1*\ra}) -- ({9*\ra-\arr},{0*\ra});
\draw[->,gray] ({8*\ra+\arr},{-1*\ra}) -- ({9*\ra-\arr},{-1*\ra});
\draw[->,gray] ({9*\ra+\arr},{1*\ra}) -- ({10*\ra-\arr},{1*\ra});
\draw[->,gray] ({9*\ra+\arr},{0*\ra}) -- ({10*\ra-\arr},{1*\ra});
\draw[->,gray] ({9*\ra+\arr},{0*\ra}) -- ({10*\ra-\arr},{0*\ra});
\draw[->,gray] ({9*\ra+\arr},{-1*\ra}) -- ({10*\ra-\arr},{0*\ra});
\draw[->,gray] ({9*\ra+\arr},{-1*\ra}) -- ({10*\ra-\arr},{-1*\ra});
\draw[->,gray] ({10*\ra+\arr},{1*\ra}) -- ({11*\ra-\arr},{.5*\ra});
\draw[->,gray] ({10*\ra+\arr},{0*\ra}) -- ({11*\ra-\arr},{.5*\ra});
\draw[->,gray] ({10*\ra+\arr},{0*\ra}) -- ({11*\ra-\arr},{-.5*\ra});
\draw[->,gray] ({10*\ra+\arr},{-1*\ra}) -- ({11*\ra-\arr},{-.5*\ra});
\draw[->,gray] ({11*\ra+\arr},{.5*\ra}) -- ({12*\ra-\arr},{.5*\ra});
\draw[->,gray] ({11*\ra+\arr},{.5*\ra}) -- ({12*\ra-\arr},{-.5*\ra});
\draw[->,gray] ({11*\ra+\arr},{-.5*\ra}) -- ({12*\ra-\arr},{-.5*\ra});
\draw[->,gray] ({12*\ra+\arr},{.5*\ra}) -- ({13*\ra-\arr},{0*\ra});
\draw[->,gray] ({12*\ra+\arr},{-.5*\ra}) -- ({13*\ra-\arr},{0*\ra});
\draw[->,gray] ({13*\ra+\arr},{0*\ra}) -- ({14*\ra-\arr},{0*\ra});
\draw[->,gray] ({14*\ra+\arr},{0*\ra}) -- ({15*\ra-\arr},{0*\ra});

\draw[-,dashed,gray] ({0*\ra},{1.5*\ra}) -- ({0*\ra},{-1.5*\ra}) node[below,text=red] {$0$};
\draw[-,dashed,gray] ({1*\ra},{1.5*\ra}) -- ({1*\ra},{-1.5*\ra}) node[below,text=red] {$1$};
\draw[-,dashed,gray] ({2*\ra},{1.5*\ra}) -- ({2*\ra},{-1.5*\ra}) node[below,text=red] {$2$};
\draw[-,dashed,gray] ({3*\ra},{1.5*\ra}) -- ({3*\ra},{-1.5*\ra}) node[below,text=red] {$3$};
\draw[-,dashed,gray] ({4*\ra},{1.5*\ra}) -- ({4*\ra},{-1.5*\ra}) node[below,text=red] {$4$};
\draw[-,dashed,gray] ({5*\ra},{1.5*\ra}) -- ({5*\ra},{-1.5*\ra}) node[below,text=red] {$5$};
\draw[-,dashed,gray] ({6*\ra},{1.5*\ra}) -- ({6*\ra},{-1.5*\ra}) node[below,text=red] {$6$};
\draw[-,dashed,gray] ({7*\ra},{1.5*\ra}) -- ({7*\ra},{-1.5*\ra}) node[below,text=red] {$7$};
\draw[-,dashed,gray] ({8*\ra},{1.5*\ra}) -- ({8*\ra},{-1.5*\ra}) node[below,text=red] {$8$};
\draw[-,dashed,gray] ({9*\ra},{1.5*\ra}) -- ({9*\ra},{-1.5*\ra}) node[below,text=red] {$9$};
\draw[-,dashed,gray] ({10*\ra},{1.5*\ra}) -- ({10*\ra},{-1.5*\ra}) node[below,text=red] {$10$};
\draw[-,dashed,gray] ({11*\ra},{1.5*\ra}) -- ({11*\ra},{-1.5*\ra}) node[below,text=red] {$11$};
\draw[-,dashed,gray] ({12*\ra},{1.5*\ra}) -- ({12*\ra},{-1.5*\ra}) node[below,text=red] {$12$};
\draw[-,dashed,gray] ({13*\ra},{1.5*\ra}) -- ({13*\ra},{-1.5*\ra}) node[below,text=red] {$13$};
\draw[-,dashed,gray] ({14*\ra},{1.5*\ra}) -- ({14*\ra},{-1.5*\ra}) node[below,text=red] {$14$};
\draw[-,dashed,gray] ({15*\ra},{1.5*\ra}) -- ({15*\ra},{-1.5*\ra}) node[below,text=red] {$15$};

\end{tikzpicture}}
	\end{center}
	\caption{Hasse diagram of UPO for $N=5$, with logistic weights (LW) of patterns in red and iLW in blue.}
	\label{fig:hasse}
\end{figure*}
\subsection{ORBGRAND}
The GRAND algorithm works as follows. 
Given a received vector $y$, hard decision is applied to its elements, obtaining an estimation $\hat{y}$, namely a vector of $N$ bits representing the most probable codeword candidate given $y$.
Then, a check function is applied to $\hat{y}$ to control if it is included in the codebook $\mathcal{C}$, thus checking if $\hat{y}$ is actually a codeword. 
For block codes, a straightforward check function is given by the parity check matrix of the code; however a given code may have a more suitable or simpler check function. 
In the case of check failure, an error pattern $e$, namely a vector of $N$ bits, is calculated, and the vector $\hat{y}+e$ is given to the check function to control if it belongs to the codebook. 
The procedure ends when a codeword is found.

Error patterns should be checked from the most probable to the least probable one. 
The scheduling of the error patterns, i.e. their order, is at the core of the different versions of GRAND.  While for binary symmetric channels it is sufficient to follow an order based on the Hamming weight of the patterns, this strategy is not correct for more complex channel models like AWGN. 
In the latter case, sorting error patterns in reliability order is in general a complex task, given the large number of patterns involved. 

The ORBGRAND algorithm \cite{ORBGRAND_first} introduces two main techniques to reduce the number of checks to be performed in order to find a candidate code word. 
To begin with, the soft received symbols of $y$ are sorted in ascending reliability order, and we call $\pi$ the resulting index permutation; if these symbols represent the log-likelihood ratios (LLRs) of the received signal, this is equivalent to sorting their absolute values in ascending order. 
Next, the error patterns are scheduled to be checked in \emph{logistic weight order} (LWO). 
The \emph{logistic weight} of an error pattern $e$ is given by $LW(e) = e \cdot i_N$, where $i_N$ is the vector of integers from 1 to $N$ and the scalar multiplication is operated over $\mathbb{N}$. 
Error patterns with the same logistic weight are checked in random order. 
However, before applying the error pattern to $\hat{y}$, its indices are sorted according to the inverse of $\pi$. 
In practice, the check function is applied to the vector $\hat{y}+\pi^{-1}(e)$.  

\subsection{BCH and polar codes}
Bose-Chaudhuri-Hocquenghem (BCH) codes \cite{BCH_forney} are block codes commonly used in communications, storage and cryptography applications. 
They can be described in the language of the theory of finite fields through the least common multiple of minimal polynomials over the field. 
One of their important features is their flexibility in terms of minimum distance, and hence error correction, given in the code design phase. 
They have simple check functions based on polynomial evaluations, making them good candidates for GRAND-based decoding. 
In some widespread cases like the two-error-correcting BCH, this check function can be implemented with a small number of syndrome calculations.

Polar codes \cite{polar} are binary block codes based on the polarization property of their kernel matrix. 
%
The involutory property of the channel transformation matrix makes these codes suitable for GRAND decoding. 
In fact, polar encoding can be efficiently performed in $\log_2N$ steps \cite{polar}, and such an encoder can be used also to check if a bit string belongs to the code book. 
A bit string check can hence be executed by encoding it and checking if the entries of the output belonging to the frozen set are all zero. 
Concatenation with a CRC, a common feature in 5G polar codes \cite{polar_5G}, allows to perform an additional check for improved precision.

\subsection{Posets}
The main problem in the implementation of a GRAND-based decoder is to sort the error patterns in reliability order, namely from the most probable to the least probable one. 
A solution to reduce the complexity of this ordering step is given by partially ordered sets. 
A partially ordered set (or poset) is a generalization of an ordering where the order relation is not connected, namely when two elements of the set are not always comparable. 
The Hasse diagram of a poset, as shown in Fig.~\ref{fig:hasse}, is a directed acyclic graph providing a graphical representation of the relations among the elements of the set. 
Each node represents an element of the set, and nodes connected through a direct path are comparable. 
Distinct nodes on the same vertical level are incomparable with each other, and so are any unconnected nodes of the graph. 

The Logistic Weight Order (LWO) used in \cite{ORBGRAND_first} is an example of poset for the set of the error patterns, and it is defined by the logistic weight relation, namely $a <_{LWO} b$ iff $LW(a) < LW(b)$. 
This poset can be described as the set of partitions into distinct summands, namely the $i$-th vertical level includes all the possible decomposition of integer $i$ as the sum of integers, where each summand appears only once. 
Decompositions are expressed as binary vectors where index $j$ is set to 1 if integer $j+1$ is included among the summands. 

\section{ORBGRAND sequence generation}
In this Section, we analyze the structure of error patterns and identify a universal partial order that is valid for any channel. 
Based on this partial order, we will show that the LWO is well defined and propose a sequential algorithm to generate the LWO sequence: the computation of the next error pattern requires only the knowledge of the previous error pattern and its logistic weight.
Moreover, we propose a novel ordering of error patterns that gives priority to low Hamming weight patterns and provides better performance compared to LWO.

\subsection{Universal partial order}
The Universal Partial Order (UPO) is a poset for strings of bits. 
This ordering has been introduced in \cite{UPO_construction} to reduce the design complexity of polar codes by avoiding comparisons between virtual channels that are ordered through the partial order. 
This poset is generated by two rules, applied to strings of bits of length $N$. 
Given two error patterns $e^1=(e^1_0,\dots,e^1_{N-1})$ and $e^2=(e^2_0,\dots,e^2_{N-1})$, where $\mathbb{P}(e^a_i=1) \geq \mathbb{P}(e^a_{i+1}=1)$, the two basic rules for the UPO are: 
\begin{itemize}
\item \textbf{Addition rule:} if the two strings differ only in entry $t$ and $e^1_{t}=0$ while $e^2_{t}=1$, then $e^1 <_{UPO} e^2$;
\item \textbf{Right swap rule:} if the two strings differ only in two consecutive entries $t$ and $t+1$, where and $e^1_{t}e^1_{t+1}=10$ while $e^2_{t}e^2_{t+1}=01$, then $e^1 <_{UPO} e^2$.
\end{itemize}
These rules can be iterated to create a chain of comparisons among error patterns. 
The Hasse diagram of this relation is depicted in Figure~\ref{fig:hasse} for $N=5$.

It is quite obvious that the UPO is valid for the ORBGRAND algorithm, where bit indices are sorted in descending reliability order and thus the condition $\mathbb{P}(e^a_i=1) \geq \mathbb{P}(e^a_{i+1}=1)$ is fulfilled.
This partial order is universal in the sense that it is independent from the channel used. 
Every error pattern ordering under ORBGRAND must follow the UPO rules in order to be well defined. 
As an example, the LWO ordering proposed in \cite{ORBGRAND_first} is compliant with the UPO: a level of the Hasse diagram of the UPO contains all the patterns of the same logistic weight \cite{min_networks_ranking}.
In fact, an error pattern $e$ has logistic weight $l$ if and only if the sum of the indices of the non-zero entries of $e$ is equal to $l$; this property is highlighted in Figure~\ref{fig:hasse}. 
Since these patterns cannot be compared using the partial order, these nodes of the diagram are not connected by edges, and hence cannot be sorted according the UPO. 

Based on this parallelism between UPO and LWO, we propose a novel sequential method to calculate the LWO error pattern sequence by following the Hasse diagram of the UPO poset in the order of the levels. 
Our method allows for a sequential computation of the error patterns using only the previously calculated one, and so is suitable in scenarios where memory consumption is an issue and the sequence should not be pre-computed and stored. 
The idea is to sort all error patterns of a given logistic weight $h$ from the smallest Hamming weight (described in Algorithm~\ref{algo_MIP}) to the one having the largest Hamming weight (described in Algorithm~\ref{algo_islast}). 
The full method is described in Algorithm~\ref{algo_hasse}. 
In practice, if the last error pattern is not the one with the largest Hamming weight having its logistic weight, the next error pattern is calculated by finding the leftmost one of the error pattern (excluding the first two entries) and substituting it with the sub-pattern having the smallest Hamming weight possible. 
This process ends when the all-ones error pattern is reached or may be early terminated when a certain number of error patterns has been calculated. 

Algorithm~\ref{algo_hasse} is used in the remainder of this paper as specific implementation to check following the LWO sequence.

\begin{algorithm}
\caption{$NextErrorPattern(errorPattern)$} \label{algo_hasse}
\begin{algorithmic}[1]
\STATE $errorPattern$: previous error pattern
\STATE $lw = LogisticWeight(errorPattern)$
\STATE $N = length(errorPattern)$
\IF{$IsLast(errorPattern)$} 
	\RETURN $MaxIntegerPartition(N,lw+1)$
\ELSE
	\WHILE{$I(1==1)$}
		\STATE $lm = LeftmostOne(errorPattern(3:N))+2$
		\STATE $r = LogisticWeight(errorPattern(lm+1:N))$	 
		\IF{$\frac{lm(lm-1)}{2} + r < lw$}
			\STATE $errorPattern(1:lm) = 0$	
		\ELSE
			\RETURN $[MaxIntegerPartition(lw-r,lm-1) , 0 , errorPattern(lm+1:N)]$	
		\ENDIF	
	\ENDWHILE	
\ENDIF	
\end{algorithmic}
\end{algorithm}
\begin{algorithm}
\caption{$IsLast(errorPattern)$} \label{algo_islast}
\begin{algorithmic}[1]
\STATE $errorPattern$: last error pattern
\STATE $N = Length(errorPattern)$
\STATE $k = LogisticWeight(errorPattern)$
\IF{$k==0$} 
	\RETURN $true$
\ENDIF	
\FOR{$i=N,i--,1$}
	\IF{$i(i-1) < 2k$}
		\IF{$errorPattern(i)==0$}
			\RETURN $false$
		\ENDIF 
		\STATE $k=k-i$
		\IF{$k==0$}
			\RETURN $true$
		\ENDIF 
	\ENDIF 
\ENDFOR	
\end{algorithmic}
\end{algorithm}
\begin{algorithm}
\caption{$MaxIntegerPartition(N,k)$} \label{algo_MIP}
\begin{algorithmic}[1]
\STATE $N$: error pattern length
\STATE $k$: desired logistic weight
\STATE $errorPattern = zeros(1,N)$
\FOR{$i=N,i--,1$}
	\IF{$i \leq k$}
		\STATE $errorPattern(i)=1$
		\STATE $k=k-i$
		\IF{$k==0$}
			\RETURN $errorPattern$
		\ENDIF 
	\ENDIF 
\ENDFOR
\RETURN $Error()$
\end{algorithmic}
\end{algorithm}
%

\subsection{Improved error pattern sequence} 
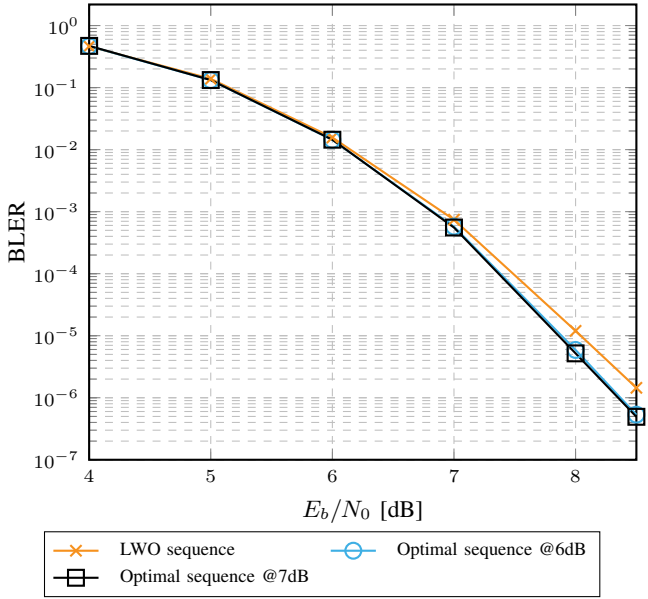
\begin{figure}[t!]
    \centering
		  \begin{tikzpicture}
  \pgfplotsset{
    label style = {font=\fontsize{9pt}{7.2}\selectfont},
    tick label style = {font=\fontsize{7pt}{7.2}\selectfont}
  }

\begin{axis}[
	scale = 1,
    ymode=log,
    xlabel={$E_b/N_0$ [\text{dB}]}, xlabel style={yshift=0.4em},
    ylabel={BLER}, ylabel style={yshift=-0.75em},
    grid=both,
    ymajorgrids=true,
    xmajorgrids=true,
    grid style=dashed,
    mark options=solid,
    width=1\columnwidth,
    thick,
        xmin=4,
        xmax=8.5,
        ymin=1e-7,
    mark size=3,
    legend style={
      anchor={center},
      cells={anchor=west},
      mark options=solid,
      column sep= 2mm,
      font=\fontsize{7pt}{7.2}\selectfont,
    },
    legend to name=BLER_ORB_opt,
    legend columns=2,
]

\addplot[
    color=BurntOrange,
    mark=x,
    thick,
    mark size=3,
]
table {
1.0 0.997
2.0 0.973
3.0 0.838
4.0 0.46296
5.0 0.13858
6.0 0.015695
7.0 0.00074333
8.0 1.198e-005
8.5 1.4458e-006
};
\addlegendentry{LWO sequence}


\addplot[
    color=CornflowerBlue,
    mark=o,
    thick,
    mark size=3,
]
table {
1.0 0.997
2.0 0.972
3.0 0.839
4.0 0.46685
5.0 0.132
6.0 0.014374
7.0 0.00057448
8.0 5.8962e-006
8.5 5.3334e-007
};
\addlegendentry{Optimal sequence @6dB}

\addplot[
    color=black,
    mark=square,
    thick,
    mark size=3,
]
table {
1.0 0.997
2.0 0.972
3.0 0.841
4.0 0.47081
5.0 0.13242
6.0 0.01446
7.0 0.00055507
8.0 5.1954e-006
8.5 4.93102e-007
};
\addlegendentry{Optimal sequence @7dB}

%

\end{axis}
\end{tikzpicture}
    \ref{BLER_ORB_opt}
    \caption{BLER of BCH(127,113,2) with ORBGRAND, $Q=35$.}
    \label{fig:orb_reorb}
\end{figure}

The LWO sequence of error patterns proposed for ORBGRAND in \cite{ORBGRAND_first} has been shown to be implementable in hardware with reasonable complexity \cite{ORBGRAND_arch}.
However, it can be observed that this sequence is suboptimal, especially as the signal-to-noise ratio (SNR) increases.
In fact, at high SNR, error patterns with Hamming weight $h$ larger than 1 are less likely to lead to successful decoding.
Figure \ref{fig:orb_reorb} shows the BLER performance of ORBGRAND with a maximum number of patterns $Q=35$ for a BCH code with $N=127$, $K=113$, and 2-error correction capability. 
The LWO sequence in Figure \ref{fig:orb_reorb} is compared to two other sequences of $Q$ patterns empirically observed at $E_b/N_0$=6dB and 7dB. 
Even with such a small pattern set, the LWO sequence suffers from noticeable loss at high SNR.  
As expected, an optimal sequence should favor low Hamming weight over low logistic weight.

An error pattern sequence based on empirical observation of the error pattern distribution is however difficult to implement, since it requires storage of a potentially very high number of patterns, as they cannot be algorithmically created on-the-fly. 
We thus propose a new partial order based on LWO, but giving more importance to the Hamming weight of error patterns. 
This \emph{improved logistic weight order} (iLWO) computes the weight of an error pattern $e=(e_0,\dots,e_{N-1})$ as follows: if $j = (j_0,\dots,j_{h-1})$ represents the ordered vector of the indices of the nonzero entries of $e$, with $h$ denoting its length, then
\begin{equation}
iLW(e) = \sum_{i=0}^{h-1} (i+1)\cdot(j_i+1)~.
\end{equation}
This error pattern weight penalizes patterns with high Hamming weight $h$ by assigning growing cost to additional flipped bits. 
In practice, the weight of each entry of the error pattern is given by its position, like for LWO, multiplied by the number of nonzero entries at its left plus one, namely on its partial Hamming weight. 
As an example, consider the sequence $e = 01100$ from Fig.~\ref{fig:hasse}: we have $LW(e) = 2+3=5$, while $iLW(e) = 2 + 2 \cdot 3 = 8$.
Note that the proposed iLWO is compliant with UPO: the two basic UPO rules can easily be verified to hold for the iLWO. 
In Figure~\ref{fig:hasse}, the iLWs of error patterns of length $N=5$ are indicated in blue. 

\section{Performance Evaluation} 
%
%
%
In this section we evaluate the performance of the proposed iLWO sequence compared to the state-of-the-art sequence generator methods for ORBGRAND decoding of BCH and polar codes.
A first general observation can be made about GRAND-based decoding algorithms, of particular interest when applied to BCH codes.
For codes having minimum distance $2t+1$, considering patterns with Hamming weight $h\ge 2t+1$ will not be very beneficial to the error-correction performance of the algorithm, as it will lead to miscorrections, especially when the number of errors introduced by the channel in each codeword is close to 1.
For this reason, we have observed that the performance of the considered algorithms when the error patterns are limited to a maximum $h$=4 and $h$=5 is very similar, as $2t+1=5$ for BCH(127,113,2).

\begin{figure}[t!]
    \centering
		  \begin{tikzpicture}
  \pgfplotsset{
    label style = {font=\fontsize{9pt}{7.2}\selectfont},
    tick label style = {font=\fontsize{7pt}{7.2}\selectfont}
  }

\begin{axis}[
	scale = 1,
    ymode=log,
    xlabel={$E_b/N_0$ [\text{dB}]}, xlabel style={yshift=0.4em},
    ylabel={BLER}, ylabel style={yshift=-0.75em},
    grid=both,
    ymajorgrids=true,
    xmajorgrids=true,
    grid style=dashed,
    mark options=solid,
    width=1\columnwidth,
    thick,
        xmin=4,
        xmax=8,
        ymin=1e-7,
    mark size=3,
    legend style={
      anchor={center},
      cells={anchor=west},
      mark options=solid,
      column sep= 2mm,
      font=\fontsize{7pt}{7.2}\selectfont,
    },
    legend to name=BLER_ORB_opt_5,
    legend columns=2,
]

\addplot[
    color=BurntOrange,
    mark=x,
    thick,
    mark size=3,
]
table {
1           0.994   
2           0.958   
3           0.758   
4           0.346   
5        0.078125   
6      0.00559722   
7     0.000274305   
8    3.37869e-006   
};
\addlegendentry{LWO sequence, $Q=10^2$}

\addplot[
    color=BurntOrange,
    mark=o,
    thick,
    mark size=3,
]
table {
1           0.995    
2            0.96    
3           0.765    
4           0.366    
5       0.0919118    
6       0.0059312    
7     0.000135009    
8    5.53916e-007  
};
\addlegendentry{iLWO sequence, $Q=10^2$}

\addplot[
    color=CornflowerBlue,
    mark=x,
    thick,
    mark size=3,
]
table {
1            0.98     
2           0.884     
3           0.545     
4           0.155     
5       0.0146007     
6     0.000828988     
7    2.50511e-005     
8    3.12822e-007
};
\addlegendentry{LWO sequence, $Q=10^3$}
\addplot[
    color=CornflowerBlue,
    mark=o,
    thick,
    mark size=3,
]
table {
1           0.979   
2           0.888   
3            0.57   
4            0.18   
5       0.0175101   
6     0.000499214   
7    3.40646e-006   
};
\addlegendentry{iLWO sequence, $Q=10^3$}

\addplot[
    color=black,
    mark=x,
    thick,
    mark size=3,
]
table {
1           0.956   
2           0.776   
3           0.357   
4       0.0769231   
5      0.00599664   
6      0.00035306   
7    7.50279e-006   
};
\addlegendentry{LWO sequence, $Q=10^4$}

\addplot[
    color=black,
    mark=o,
    thick,
    mark size=3,
]
table {
1           0.965   
2             0.8   
3           0.379   
4       0.0807754   
5       0.0048506   
6    9.76108e-005   
7    7.45934e-007   
};
\addlegendentry{iLWO sequence, $Q=10^4$}

\end{axis}
\end{tikzpicture}
    \ref{BLER_ORB_opt_5}
    \caption{BLER for BCH(127,113,2), no limit on $h$.}
    \label{fig:orb_ieorb_nolim}
\end{figure}
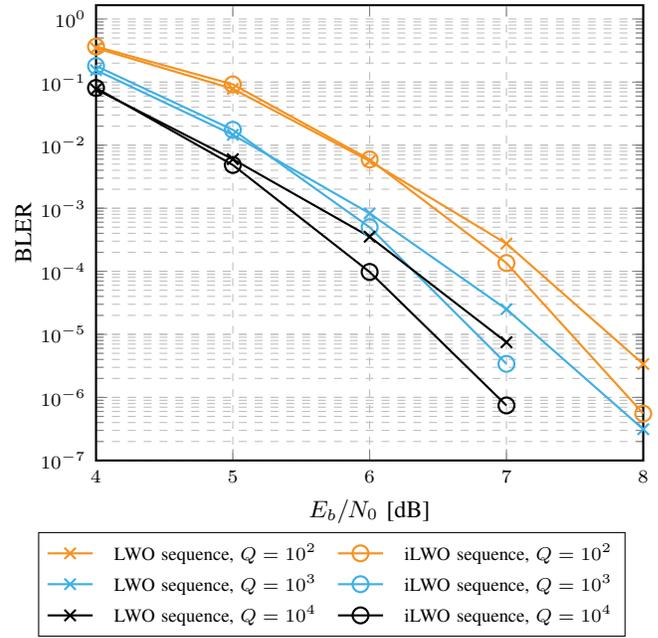
\begin{figure}[t!]
    \centering
		  \begin{tikzpicture}
  \pgfplotsset{
    label style = {font=\fontsize{9pt}{7.2}\selectfont},
    tick label style = {font=\fontsize{7pt}{7.2}\selectfont}
  }

\begin{axis}[
	scale = 1,
    ymode=log,
    xlabel={$E_b/N_0$ [\text{dB}]}, xlabel style={yshift=0.4em},
    ylabel={BLER}, ylabel style={yshift=-0.75em},
    grid=both,
    ymajorgrids=true,
    xmajorgrids=true,
    grid style=dashed,
    mark options=solid,
    width=1\columnwidth,
    thick,
        xmin=4,
        xmax=8,
        ymin=1e-7,
    mark size=3,
    legend style={
      anchor={center},
      cells={anchor=west},
      mark options=solid,
      column sep= 2mm,
      font=\fontsize{7pt}{7.2}\selectfont,
    },
    legend to name=BLER_ORB_opt_5,
    legend columns=2,
]

\addplot[
    color=BurntOrange,
    mark=x,
    thick,
    mark size=3,
]
table {
1           0.999
2           0.939
3            0.75
4           0.348
5        0.071582
6      0.00525873
7     0.000207691
8    3.63767e-006
};
\addlegendentry{LWO sequence, $Q=10^2$}

\addplot[
    color=BurntOrange,
    mark=o,
    thick,
    mark size=3,
]
table {
1           0.998
2           0.943
3           0.771
4            0.35
5       0.0881834
6      0.00561735
7     0.000122338
8    1.22824e-006
};
\addlegendentry{iLWO sequence, $Q=10^2$}

\addplot[
    color=CornflowerBlue,
    mark=x,
    thick,
    mark size=3,
]
table {
1           0.983
2            0.87
3           0.548
4           0.166
5       0.0155594
6     0.000728401
7    2.25856e-005
8	 5.41228e-007
};
\addlegendentry{LWO sequence, $Q=10^3$}
\addplot[
    color=CornflowerBlue,
    mark=o,
    thick,
    mark size=3,
]
table {
1           0.987
2           0.874
3           0.557
4           0.177
5       0.0160565
6     0.000527123
7    7.36995e-006
8	 1.03e-007  
};
\addlegendentry{iLWO sequence, $Q=10^3$}

\addplot[
    color=black,
    mark=x,
    thick,
    mark size=3,
]
table {
1           0.971
2           0.764
3           0.361
4        0.085034
5      0.00573888
6      0.00031757
7     9.9459e-006
8	  2.01532e-007 
};
\addlegendentry{LWO sequence, $Q=10^4$}

\addplot[
    color=black,
    mark=o,
    thick,
    mark size=3,
]
table {
1           0.969
2           0.785
3           0.397
4       0.0839631
5      0.00597122
6     0.000195512
7     4.7376e-006
8	  7.1563e-008
};
\addlegendentry{iLWO sequence, $Q=10^4$}

\end{axis}
\end{tikzpicture}
    \ref{BLER_ORB_opt_5}
    \caption{BLER for PC(128,114), 6-bit CRC, no limit on $h$.}
    \label{fig:orb_ieorb_nolim_polar}
\end{figure}
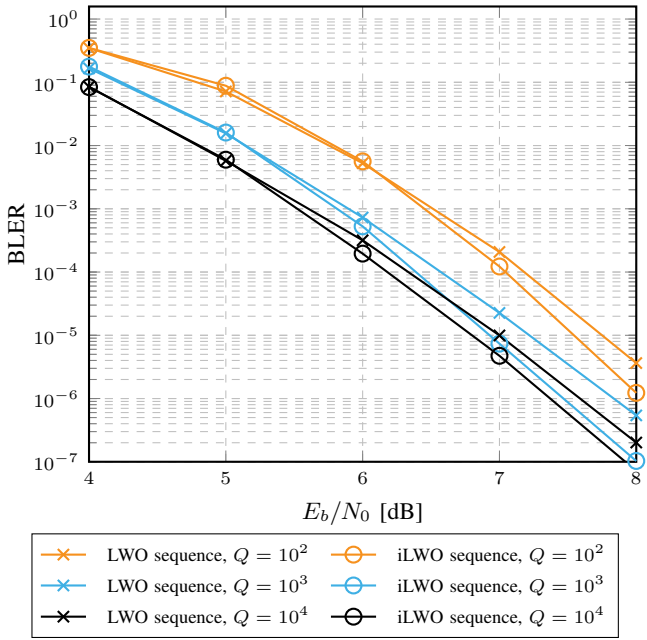

To begin with, Figure~\ref{fig:orb_ieorb_nolim} shows the BLER performance of ORBGRAND decoding of a BCH(127,113,2) code for different values of maximum number of patterns to check $Q$. 
Simulation results show performance improvement for iLWO up to 0.5dB at BLER=$10^{-5}$ compared to the original LWO, with increasing gains as the SNR rises and the BLER decreases. 
Moreover, iLWO can match the performance of LWO by reducing $Q$ of an order of magnitude; in fact, iLWO with $Q=10^3$ outperforms LWO with $Q=10^4$  at $E_b/N_0\geq$ 6.5dB.
This is due to the fact that the LWO sequence tends to assign a disproportionately low weight to patterns with high Hamming weight, compared to their likelihood of occurrence, and thus incurs BLER degradation with respect to iLWO.
The same trends are observed in Figure~\ref{fig:orb_ieorb_nolim_polar}, where the performance of ORBGRAND with LWO and iLWO is evaluated on a polar code taken from the 5G standard, with $N=128$ and $K=114$, and a CRC of 6 bits.

\begin{figure}[t!]
    \centering
		  \begin{tikzpicture}
  \pgfplotsset{
    label style = {font=\fontsize{9pt}{7.2}\selectfont},
    tick label style = {font=\fontsize{7pt}{7.2}\selectfont}
  }

\begin{axis}[
	scale = 1,
    ymode=log,
    xlabel={$E_b/N_0$ [\text{dB}]}, xlabel style={yshift=0.4em},
    ylabel={BLER}, ylabel style={yshift=-0.75em},
    grid=both,
    ymajorgrids=true,
    xmajorgrids=true,
    grid style=dashed,
    mark options=solid,
    width=1\columnwidth,
    thick,
        xmin=4,
        xmax=8,
        ymin=1e-7,
    mark size=3,
    legend style={
      anchor={center},
      cells={anchor=west},
      mark options=solid,
      column sep= 2mm,
      font=\fontsize{7pt}{7.2}\selectfont,
    },
    legend to name=BLER_ORB_opt_4,
    legend columns=2,
]

\addplot[
    color=BurntOrange,
    mark=x,
    thick,
    mark size=3,
]
table {
1           0.994  
2           0.958  
3           0.758  
4           0.346  
5        0.078125  
6      0.00559722  
7     0.000274305  
8    3.37869e-006  
};
\addlegendentry{LWO sequence, $Q=10^2$}

\addplot[
    color=BurntOrange,
    mark=o,
    thick,
    mark size=3,
]
table {
1           0.995  
2            0.96  
3           0.765  
4           0.366  
5       0.0919118  
6       0.0059312  
7     0.000135009  
8    5.53916e-007  
};
\addlegendentry{iLWO sequence, $Q=10^2$}

\addplot[
    color=CornflowerBlue,
    mark=x,
    thick,
    mark size=3,
]
table {
1           0.981  
2            0.89  
3            0.55  
4            0.17  
5       0.0176897  
6     0.000769693  
7    2.14646e-005  
8    2.70385e-007  
};
\addlegendentry{LWO sequence, $Q=10^3$}
\addplot[
    color=CornflowerBlue,
    mark=o,
    thick,
    mark size=3,
]
table {
1            0.98 
2           0.889 
3           0.575 
4           0.183 
5       0.0175101 
6     0.000545756 
7    3.82231e-006 
};
\addlegendentry{iLWO sequence, $Q=10^3$}

\addplot[
    color=black,
    mark=x,
    thick,
    mark size=3,
]
table {
1           0.969 
2           0.817 
3           0.415 
4           0.108 
5       0.0083167 
6     0.000312992 
7    4.79786e-006 
};
\addlegendentry{LWO sequence, $Q=10^4$}

\addplot[
    color=black,
    mark=o,
    thick,
    mark size=3,
]
table {
1           0.969 
2           0.826 
3           0.424 
4           0.111 
5      0.00870928 
6     0.000242982 
7    1.46451e-006 
};
\addlegendentry{iLWO sequence, $Q=10^4$}

\end{axis}
\end{tikzpicture}
    \ref{BLER_ORB_opt_4}
    \caption{BLER for BCH(127,113,2), maximum $h=4$.}
    \label{fig:orb_ieorb4}
\end{figure}
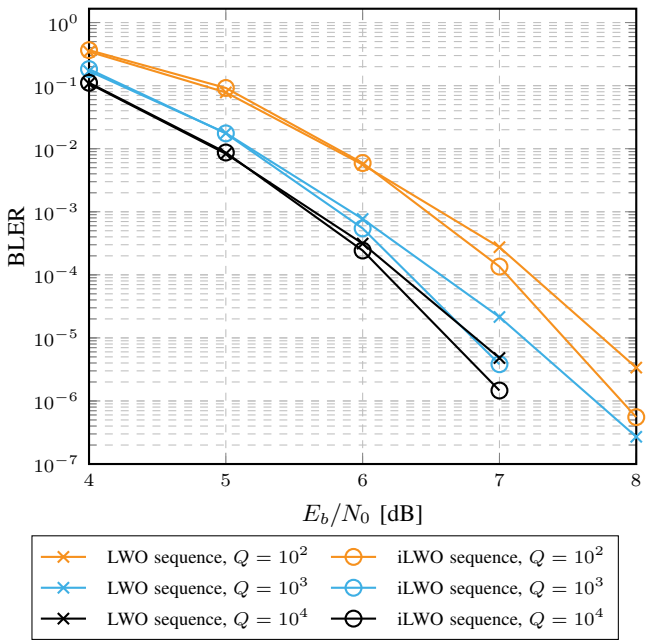
Next, in Figure~\ref{fig:orb_ieorb4} we limit the maximum Hamming weight of the candidate error patterns to $h=4$.
This limitation is useful for shift-register-based hardware architectures, similar to that proposed for ORBGRAND in \cite{ORBGRAND_arch}. 
As expected given the minimum distance of the BCH(127,113,2) code, the performance of ORBGRAND with $h\leq$4 is similar to the unlimited case. 
Moreover, if this limit is raised to 5, the performance is equivalent to the unlimited case; higher $h$ have thus been omitted from the paper.  

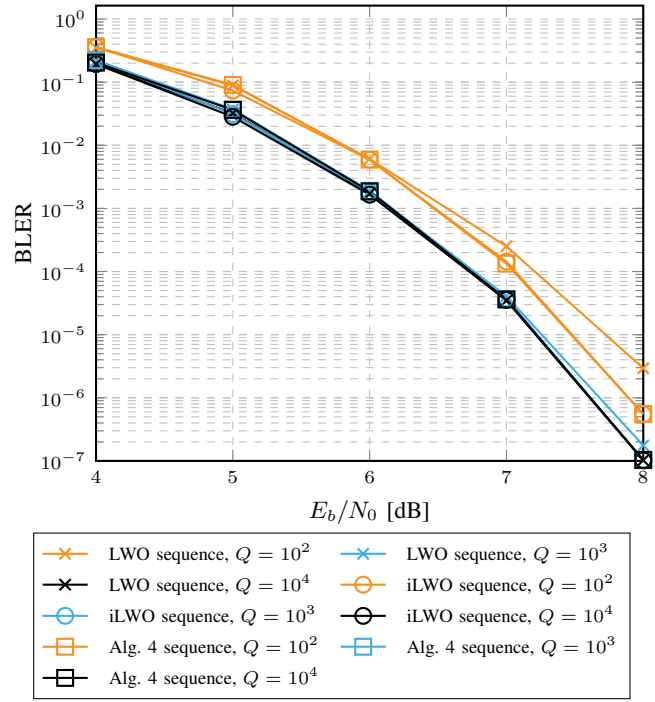
\begin{figure}[t!]
    \centering
		  \begin{tikzpicture}
  \pgfplotsset{
    label style = {font=\fontsize{9pt}{7.2}\selectfont},
    tick label style = {font=\fontsize{7pt}{7.2}\selectfont}
  }

\begin{axis}[
	scale = 1,
    ymode=log,
    xlabel={$E_b/N_0$ [\text{dB}]}, xlabel style={yshift=0.4em},
    ylabel={BLER}, ylabel style={yshift=-0.75em},
    grid=both,
    ymajorgrids=true,
    xmajorgrids=true,
    grid style=dashed,
    mark options=solid,
    width=1\columnwidth,
    thick,
        xmin=4,
        xmax=8,
        ymin=1e-7,
    mark size=3,
    legend style={
      anchor={center},
      cells={anchor=west},
      mark options=solid,
      column sep= 2mm,
      font=\fontsize{7pt}{7.2}\selectfont,
    },
    legend to name=BLER_APPROX,
    legend columns=2,
]

\addplot[
    color=BurntOrange,
    mark=x,
    thick,
    mark size=3,
]
table {
1           0.994 
2           0.956 
3           0.756 
4           0.349 
5        0.088968 
6      0.00606281 
7     0.000251144 
8    2.93786e-006 
};
\addlegendentry{LWO sequence, $Q=10^2$}

\addplot[
    color=CornflowerBlue,
    mark=x,
    thick,
    mark size=3,
]
table {
1           0.981
2             0.9
3           0.614
4           0.219
5       0.0369004
6      0.00186019
7    4.06743e-005
8	 1.75827e-007
};
\addlegendentry{LWO sequence, $Q=10^3$}

\addplot[
    color=black,
    mark=x,
    thick,
    mark size=3,
]
table {
1           0.976
2            0.87
3           0.554
4           0.193
5       0.0322061
6      0.00171086
7    3.46577e-005
8    1.02392e-007
};
\addlegendentry{LWO sequence, $Q=10^4$}

\addplot[
    color=BurntOrange,
    mark=o,
    thick,
    mark size=3,
]
table {
1           0.995
2            0.96
3           0.765
4         0.36049
5       0.0734646
6      0.00590138
7     0.000143824
8     5.4867e-007
};
\addlegendentry{iLWO sequence, $Q=10^2$}

\addplot[
    color=CornflowerBlue,
    mark=o,
    thick,
    mark size=3,
]
table {
1           0.979   
2           0.905   
3           0.617   
4        0.216357   
5       0.0311857   
6      0.00181222   
7    3.74021e-005   
8	 1.02342e-007	
};
\addlegendentry{iLWO sequence, $Q=10^3$}

\addplot[
    color=black,
    mark=o,
    thick,
    mark size=3,
]
table {
1           0.974   
2           0.875   
3           0.557   
4        0.193349   
5       0.0282773   
6      0.00165158   
7    3.55175e-005   
8	 1.02342e-007	
};
\addlegendentry{iLWO sequence, $Q=10^4$}

\addplot[
    color=BurntOrange,
    mark=square,
    thick,
    mark size=3,
]
table {
1           0.995 
2           0.959 
3           0.764 
4           0.359 
5       0.0904977 
6      0.00593718 
7     0.000133322 
8    5.53945e-007 
};
\addlegendentry{Alg. \ref{alg:scaledweights} sequence, $Q=10^2$}

\addplot[
    color=CornflowerBlue,
    mark=square,
    thick,
    mark size=3,
]
table {
1            0.98    
2            0.91    
3           0.621    
4           0.223    
5       0.0370645    
6      0.00190309    
7    3.67256e-005    
8    1.03657e-007 	
};
\addlegendentry{Alg. \ref{alg:scaledweights} sequence, $Q=10^3$}

\addplot[
    color=black,
    mark=square,
    thick,
    mark size=3,
]
table {
1           0.975    
2           0.895    
3           0.588    
4           0.204    
5       0.0363108    
6      0.00186171    
7    3.60912e-005    
8    1.03662e-007 	
};
\addlegendentry{Alg. \ref{alg:scaledweights} sequence, $Q=10^4$}

\end{axis}
\end{tikzpicture}
    \ref{BLER_APPROX}
    \caption{BLER for BCH(127,113,2), maximum $h=3$.}
    \label{fig:approx}
\end{figure}
Finally, we compare our proposal with state-of-the-art hardware sequence generator, namely imposing the $h$ limit to 3. 
In order to help the design of a hardware architecture for our iLWO sequence we present Algorithm \ref{alg:scaledweights}, a method to calculate a close approximation to the iLWO sequence, that can easily be run on-the-fly with low complexity.
The algorithm produces error patterns with HW~$\le3$, like the architecture in \cite{ORBGRAND_arch}, but can be easily extended to higher $h$ by incuding state machines of growing complexity. 
The algorithm output $\mathcal{E}$ the set of indices flipped by an error pattern.
All conditions in Algorithm \ref{alg:scaledweights} and \ref{alg:HW3} assume that any flipped index is $<N$: this check has been omitted in the text for clarity. 
The performance of the proposed approximate algorithm, compared to the LWO and iLWO sequences, is shown in Figure~\ref{fig:approx}. 
The sequence defined in the hardware implementation in \cite{ORBGRAND_arch} yields the same BLER as the LWO curve; since \cite{ORBGRAND_first} does not specify how to sort error patterns with the same weight, in our implementation of LWO we have sorted them in ascending order of Hamming weight, as in \cite{ORBGRAND_arch}. 
The curves show a saturation of the BLER for $Q \geq 10^3$.
This is due to the fact that limiting the maximum $h$ to 3 allows the different pattern sequences to have a large intersection as $Q$ increases. 
It can be observed that the curve obtained with Alg.~\ref{alg:scaledweights} is superimposed to that of iLWO, and thus yields the same gains with respect to LWO as the ideal iLWO.

\begin{algorithm}
\begin{algorithmic}
\STATE $dw$: Desired weight 
\STATE $h3dw$: Last weight for which $h$=3 patterns were created
\STATE $m,n$: Indices of the $2^{nd}$ and $3^{rd}$ bit of the first $h$=3 pattern created at weight $h3dw$
\STATE $\mathcal{E}=\{dw-1\}$
\IF {$dw>4$} 
\STATE $w=\lfloor (dw-1)/2 \rfloor-1$
\STATE $k=(dw-1)~mod~2$
\IF {$w\ge1$ \&\& $k>w$} 
        \STATE $\mathcal{E}=\{k,w\}$
        \WHILE{$w-1>k+2$}
	        \STATE $w=w-1$, $k=k+2$
	        \STATE $\mathcal{E}=\{k,w\}$        
        \ENDWHILE
\ENDIF
\IF {$dw==14$} 
        \STATE $l=0$, $m=1$, $n=2$
        \STATE $\mathcal{E}=\{l,m,n\}$        
        \STATE $h3dw=dw$
\ELSIF {$dw>14$} 
	    \STATE $m_{old}=m$, $n_{old}=n$
		\IF {$h3dw=dw-1$}
			\IF {$m>1$}		
        		\STATE $l=0$, $m=m-1$, $n=n+1$
	            \STATE $\mathcal{E}=\{l,m,n\}$   
    	        \STATE $h3dw=dw$  
				\STATE CreateRemainingh3Patterns$(l,m,n)$          
			\ENDIF	
			\IF {$m_{old}+2<n_{old}-1$}		
			    \STATE $l=0$, $m_1=m-1$, $n_1=n+1$
  	    		\IF {$h3dw=dw-1$}
				    \STATE $m=m_1$, $n=n_1$  	    		
  	    		\ENDIF
	            \STATE $\mathcal{E}=\{l,m_1,n_1\}$   
    	        \STATE $h3dw=dw$  
				\STATE CreateRemainingh3Patterns$(l,m_1,n_1)$          
			\ENDIF	
		\ELSIF {$h3dw=dw-2$ \&\& $n>m+1$}
        	\STATE $l=0$, $m=m+1$
            \STATE $\mathcal{E}=\{l,m,n\}$        
            \STATE $h3dw=dw$
  			\STATE CreateRemainingh3Patterns$(l,m,n)$
		\ELSIF {$h3dw=dw-3$}
			 \STATE $l=0$, $m=1$, $n=n+1$
		     \STATE $\mathcal{E}=\{l,m,n\}$        
		     \STATE $h3dw=dw$
			\STATE CreateRemainingh3Patterns$(l,m,n)$			
		\ENDIF
\ENDIF
\ENDIF
\end{algorithmic}
\caption{Approximate iLWO sequence with Hamming weight $h\le$3 for weight $dw$.}
\label{alg:scaledweights}
\end{algorithm}

\begin{algorithm}
\begin{algorithmic}
			\STATE $a=l$
            \STATE $b=m$
	        \STATE $c=n$
            \WHILE{$c-1>b+1$}
        		\STATE $a=a+1$
    		    \STATE $b=b+1$
		        \STATE $c=c-1$
                \STATE $\mathcal{E}=\{a,b,c\}$        
            \ENDWHILE
\end{algorithmic}
\caption{CreateRemainingh3Patterns$(l,m,n)$}
\label{alg:HW3}
\end{algorithm}

\section{Conclusion}
In this work, we have shown how the error pattern schedule based on logistic weight order (LWO) used by the ORBGRAND decoding algorithm is linked to the idea of universal partial order, and used such link to propose an on-the-fly, sequential error pattern generation algorithm. 
We pointed out that the LWO sequence is suboptimal, and proposed an improved schedule that yields up to 0.5dB gains at BLER=$10^{-5}$. 
Finally, we proposed a low-complexity approximated algorithm fit for hardware implementation of the new schedule, and showed that its performance follows the curve of the ideal schedule.


\end{document}